**Topology-dependent anomalous dynamics of ring and linear DNA are sensitive to cytoskeleton crosslinking**

Devynn M. Wulstein+, Kathryn E. Regan+, Jonathan Garamella, Ryan J. McGorty#, Rae M. Robertson-Anderson#*

*Department of Physics & Biophysics, University of San Diego, San Diego CA 92110*

+equal contributions
#equal contributions
*corresponding author, randerson@sandiego.edu

**Abstract**

Cytoskeletal crowding plays a key role in the diffusion of DNA molecules through the cell, acting as a barrier to effective intracellular transport and conformational stability required for such processes as transfection, viral infection, and gene therapy. Here we elucidate the transport properties and conformational dynamics of linear and ring DNA molecules diffusing through entangled and crosslinked composite networks of actin and microtubules. We couple single-molecule conformational tracking with differential dynamic microscopy to reveal that ring and linear DNA exhibit surprisingly distinct transport properties that are influenced differently by cytoskeleton crosslinking. Ring DNA coils are swollen and undergo heterogeneous and biphasic subdiffusion that is hindered by crosslinking. Conversely, crosslinking actually facilitates the single-mode subdiffusion that compacted linear chains exhibit. Our collective results demonstrate that transient threading by cytoskeleton filaments plays a key role in the dynamics of ring DNA, whereas the mobility of the cytoskeleton dictates transport of linear DNA.

**Introduction**

DNA, a ubiquitous biopolymer in eukaryotic and prokaryotic cells, occurs naturally in linear and relaxed circular (ring) topologies. Transport of these topologically-distinct biopolymers through the cytoskeleton is crucial for a wide range of processes and functions such as transcription, transformation, looping, gene expression, and gene therapy (*1–5*). However, the cytoskeleton is a crowded composite network of filamentous proteins that can restrict transport and impact the conformational stability of DNA required for these diverse processes (*6–8*). Two primary cytoskeletal proteins are semiflexible actin filaments, with a persistence length $l_p \approx 10$ μm, and rigid microtubules with $l_p \approx 1$ mm (*6, 8–10*). These biopolymers form steric entanglements with one another and are also often chemically crosslinked *via* accessory proteins to enable proliferation, differentiation, and cell migration (*8, 10–13*). The role that crosslinking plays in the viscoelastic properties of *in vitro* actin and microtubule networks has been widely studied (*10, 11, 14*). More recently, the interactions between actin and microtubules and their role in cell mechanics has begun to be explored (*15–19*). However, far less understood is the dynamics — both center-of-mass and conformational — of biopolymers like DNA existing within these composite cytoskeletal networks (*19*).

Cellular crowding has long been recognized as playing a key role in intracellular transport and conformational dynamics, and numerous studies have been devoted to understanding the complex and often anomalous diffusive properties that arise in such environments (*20–27*). However, the vast parameter space of crowding conditions and diffusing biopolymers and particles of interest has led to wide-ranging experimental results and theoretical predictions (*24, 28*). Several of these studies have reported normal

Brownian motion in which the mean-squared displacement (MSD) scales linearly with time as $MSD \sim 2Dt$ where $D$ is the diffusion coefficient (22, 27, 29, 30). Others have reported anomalous subdiffusion, in which the MSD scales as a power-law with time, $MSD \sim Kt\alpha$ where $K$ is the transport coefficient and the scaling exponent $\alpha < 1$ (19–21, 24, 25, 31–34). Crowded DNA and other polymers have also been reported to undergo compaction, swelling, or elongation depending on the topology and size of the DNA and crowders (20, 21, 35, 36). However, in the majority of these studies the crowders have been small globular proteins or synthetic polymers that cannot accurately mimic the constraints the cytoskeleton imposes.

When the crowders are sufficiently long and concentrated, as is the case for the cytoskeleton, they become entangled and their mobility is restricted. Such entangled networks impact the diffusion and conformation of tracers differently than systems of small mobile crowders. Likewise, distinct from the dynamics of spherical tracer particles, long tracer polymers like DNA embedded in an entangled network are restricted to move via curvilinear diffusion along their backbones – a process termed reptation (37, 38). However, ring polymers lack free ends required for this 'head-first' diffusive mechanism. As such, transport through entangled and crowded networks is fundamentally different for ring polymers than linear chains and is thus still a topic of great interest and debate (39–46).

Ring polymers entangled by linear chains have been predicted to assume diverse conformations that lead to multiple transport mechanisms (34, 46–53). They can be folded in half and undergo reptation-like diffusion similar to a linear chain of half the length or adopt amoeba-like conformations that diffuse similar to branched polymers. Rings can also become threaded by surrounding linear chains such that they can only diffuse by the threading chains unthreading via reptation, a process termed constraint release. This extremely slow process is essentially halted if the threading chains are crosslinked and thus cannot reptate to unthread the ring and release the constraint they impose. Simulations have shown that these multiple diffusive modes, and the interconversion between them, leads to heterogeneous transport and conformations of rings (54–56). However, experimental evidence for these varied transport modes and conformations is sparse (49, 50, 57). Further, how these idealized models translate to complex biological systems, such as DNA diffusion through the cytoskeleton, remains unknown.

Here, we couple single-molecule conformational tracking (SMCT) with differential dynamic microscopy (DDM) to characterize the dynamics of linear and ring DNA molecules crowded by entangled and crosslinked cytoskeletal networks. We find intriguing dependences of both DNA topology and cytoskeleton crosslinking on the transport and conformational dynamics of DNA over a range of spatiotemporal scales. Ring DNA exhibits biphasic subdiffusion and slow fluctuations between a broad range of swollen conformational states and corresponding transport modes. Linear DNA undergoes faster single-mode diffusion and more compact conformations with a narrow distribution of dynamical modes. Further, while crosslinking suppresses ring DNA diffusion, it enhances the diffusion of linear DNA. Finally, ensemble analysis reveals that, unlike linear DNA, rings undergo highly heterogeneous transport that cannot be fit to standard models of diffusion. These collective results suggest that threading – inaccessible to linear chains – plays a key role in the transport of ring DNA within cytoskeleton networks. Beyond the importance of our results to biological processes such as transfection, infection, and gene therapy, our work also provides key insights into the dynamics of entangled ring polymers – a topic of broad current interest (49, 54, 58–60).

## Results

As detailed in Fig. 1 and Methods, the ring and linear DNA used in our experiments have identical contour lengths of $L \cong 38$ μm (115 kbp) and topology-dependent mean end-to-end coil lengths of $R_{0,R} \cong$ 1.6 µm and $R_{0,L} \cong 2.6$ µm. The cytoskeleton networks both have a mesh size of $\xi \cong 0.81$ µm, and in the crosslinked network every filament crossing or entanglement can be assumed to be crosslinked. The longest relaxation times for entangled and crosslinked networks have been reported to be ~3.3 s and ~4.1 s, (*16*) similar to our measurement timescale for single-molecule tracking.

We use single-molecule conformational tracking to examine the effect of crosslinking on the dynamics of ring and linear DNA molecules diffusing within the described cytoskeleton networks (Fig. 1). We track the center-of-mass (COM) trajectories of an ensemble of individual molecules from which we evaluate mean-squared displacements (MSD) (Methods, Figs. 2A, S1). As shown in Fig. 2A, the MSDs are clearly sub-linear (i.e. subdiffusive), so we fit MSDs to the power-law function $MSD = Kt\alpha$, where $K$ is the transport coefficient and $\alpha$ the anomalous scaling exponent (Fig. 2B,C). As described in the Introduction, for normal Brownian motion, $\alpha = 1$ and $K=2D$, where $D$ is the diffusion coefficient.

As shown in Fig. 2, we observe a clear difference in the MSDs for ring and linear DNA in both entangled (E) and crosslinked (XL) networks. For both network architectures, ring DNA exhibits lower transport coefficients and a higher degree of subdiffusion. This behavior is in contrast to dilute conditions in which ring DNA diffuses ~1.4x faster than its linear counterpart due to its smaller conformational size (i.e. radius of gyration $R_G$) and reduced degrees of freedom (*46*). Further, while MSDs for linear DNA obey a single subdiffusive power-law ($\alpha \approx 0.73$), ring DNA MSDs display an intriguing shift from $\alpha \approx 0.65$ scaling to a more subdiffusive regime ($\alpha \approx 0.48$) at ~0.4 µm2 (Fig. 2A). These topology-dependent differences are amplified when crosslinkers are added to the networks. In fact, as shown in Fig. 2B,C, rings and linear chains exhibit *opposite* responses to crosslinking: ring DNA transport becomes slower (decrease in $K$) and more anomalous (decrease in $\alpha$), while linear DNA undergoes faster, less anomalous motion (increase in $K$ and $\alpha$). We also evaluate the distribution of time-averaged mean squared displacements of individual molecules (Fig. S1). We find that these distributions are wider for ring DNA compared to linear DNA in both network types. Further, the distribution for rings includes a fraction of nearly zero MSDs, particularly in the crosslinked network. This effect is absent in the linear DNA distributions.

To shed further light on the topology-specific differences in transport, we quantify the distributions of conformational states accessed by ring and linear DNA molecules in both network types. Specifically, as described in Methods and Fig. 1B, we calculate an effective DNA coil size, $R_{coil} = [1/2(R_{max2} + R_{min2})]_{1/2}$, and normalize by the dilute limit mean end-to-end length, $R_0 = \sqrt{6}R_G$ (*61*). Probability distributions of this reduced coil size $r_{coil}$ (Fig. 3A) as well as the mean value $<r_{coil}>$ (Fig. 3B, Table S1) and full-width-half-maximum (FWHM, Fig. 3C, Table S1), show distinct differences between the two topologies: linear DNA conformations are more compact than their dilute limit size (i.e. $r_{coil} < 1$) while ring DNA is swollen ($r_{coil} > 1$). The width of ring DNA distributions, quantified by the FWHM, is also markedly larger than for linear DNA, signifying a greater range of conformational states assumed by rings compared to linear DNA. This effect mirrors that of the transport distributions shown in Fig. S1. Crosslinking the composite networks increases $<r_{coil}>$ for both topologies, yet its impact on the range of conformations accessed (i.e. FWHM) is topology-dependent. Namely, for linear DNA there is little change in the FWHM upon crosslinking, whereas crosslinking reduces the FWHM for ring DNA.

The topology-dependent breadth in conformations, quantified by the FWHM, can arise from a heterogeneous ensemble of molecules that have different conformational states, from a homogenous ensemble in which all molecules undergo large conformational fluctuations in time, or from a combination of both. To determine which, if any, of these possibilities is dominant, we determine the extent to which molecules conformationally fluctuate or "breathe" between different conformational states. Specifically, for each tracked molecule we measure the fractional change in $R_{max}$ between varying lag times $t$, which we term the fractional fluctuation length, $L_f(t) = <|R_{max}(0) - R_{max}(t)|>/<R_{max}>$ (Fig. 3D). For the case of a heterogeneous ensemble of largely static conformations, the asymptotic value of $L_f(t)$ should be small and not correlate with the FWHM of the DNA coil size distribution, whereas for the temporally fluctuating homogenous sample, the FWHM should be proportional to the asymptotic $L_f(t)$ value. For linear DNA, $L_f(t)$ approaches a steady-state plateau value over the measurement timescale whereas ring DNA fluctuations are much slower, with $L_f(t)$ following power-law scaling over the entire measurement time. To better quantify the characteristic time and scale of conformational fluctuations, we calculate the final fractional fluctuation length $L_{f,f}$ as well as the time $\tau$ needed to reach 90% of $L_{f,f}$ (Fig. 3E, Table S1). Not only do rings fluctuate more slowly (i.e. larger $\tau$) but the scale of fluctuations $L_{f,f}$ is smaller than for linear DNA. Thus, the effective fluctuation *speed* $L_{f,f}\tau^{-1}$ for rings is 62% and 35% slower than linear DNA in entangled and crosslinked networks respectively. Given the ~2-fold larger FWHM of coil size distributions for rings compared to linear DNA (Fig. 3C), this result suggests that the distribution of conformational states for rings arises from a heterogeneous ensemble of ring molecules assuming different conformational states rather than rings undergoing enhanced conformational fluctuations in time compared to linear DNA. Further, crosslinking decreases $L_{f,f}$ and increases $\tau$ for both DNA topologies, indicating that the corresponding reduction in FWHM of rings upon crosslinking arises from suppressed fluctuations or interconversion between states rather than a reduction in the number of states the molecules access. Finally, while crosslinking amplifies the differences in COM transport dynamics between ring and linear DNA, it surprisingly gives rise to more similar conformational dynamics between topologies. Specifically, as shown in Fig. 3B-E, the topology-dependent differences between the derived quantities $<r_{coil}>$, FWHM, $L_{f,f}$, and $\tau$ are all less in crosslinked compared to entangled networks.

We show above that the larger breadth in $r_{coil}$ distributions for rings versus linear chains comes from a more heterogeneous ensemble of conformational states rather than larger conformational fluctuations in time. We argue that these conformational states, which could for example be predicted threaded, folded, and amoeba-like states for rings, are linked to different transport modes such as reptation, restricted reptation, constraint release, caged diffusion, etc. (*49, 54*). To verify this interpretation, we turn to our differential dynamic microscopy analysis (Fig. 1C, Methods), which measures the decay of density fluctuations of labeled DNA within the sample to probe the transport of large sub-ensembles of molecules over ~5x longer times than with single-molecule conformational tracking (*62*).

By evaluating the intermediate scattering functions (ISF) for all conditions, we determine the extent to which transport is heterogeneous and how long anomalous dynamics persist. As shown in Fig. 4A, ring transport is significantly slower than linear DNA, displayed as a much slower decay of the ISF and larger ISF values for all times. Additionally, while the ISF for linear DNA follows a nearly exponential decay, expected for diffusive dynamics (*19, 63*) and well-fit to standard models for subdiffusion, the ISF for ring DNA is clearly far from a simple exponential and cannot be fit to any existing models used to analyze DDM ISFs (*62–66*). This data corroborates our SMCT results that show rings exhibit much slower and more anomalous transport than linear DNA and further suggests the existence of multiple modes of transport. Further, the response of the DNA to network crosslinking is topology-dependent. Ring DNA ISFs decay more slowly in crosslinked versus entangled networks, similar to our SMCT results that show a decrease in $K$ and $\alpha$ values upon crosslinking. On the other hand, linear DNA exhibits nearly identical ISFs in the two network architectures, compared to the corresponding SMCT results that show the measured MSD and corresponding $K$ and $\alpha$ values increase upon crosslinking. Because DDM measures density fluctuations that can arise from both conformational fluctuations as well as COM transport, the *increase* in COM transport of linear DNA with crosslinking coupled with the corresponding *decrease* in conformational fluctuations may explain the similarity between the linear DNA ISFs for both network architectures. Conversely, for rings, we see a decrease in both COM transport and conformational dynamics with crosslinking which accounts for the different ISFs.

To evaluate the heterogeneity in transport modes, we consider the spread in ISFs over different spatial regions (ROIs) of the network (Fig. 4B,C). Specifically, a wider spread in the distribution of individual ISFs (grey lines) indicates more heterogeneity in transport modes. To quantify this heterogeneity, we compute the difference between the maximum and minimum ISF value for each lag time, $\Delta f(t) = f_{max}(t) - f_{min}(t)$ (Fig 4D, E). As shown, $\Delta f(t)$ for ring DNA in both network types is significantly larger than that for linear DNA. This topology-dependent difference in ISF heterogeneity is evidence that the spread measured in the conformational analysis (Fig. 3) is linked to varying transport modes, and that rings access a wider range of transport modes that are inaccessible to linear DNA. Further, we also once again see an opposite effect of crosslinking for the different topologies. Crosslinking slightly increases $\Delta f(t)$ for linear DNA while it slightly reduces $\Delta f(t)$ for rings. The reduction for rings corroborates the reduction in the FWHM of $r_{coil}$ – coupling the suppression of conformational states with that of transport modes. For linear DNA, our SMCT analysis shows a ~8% reduction in $L_f(t)$ upon crosslinking but no change in the FWHM of the $r_{coil}$ distribution, which includes both temporal fluctuations and heterogeneities. To maintain the same FWHM there must be an increase in the heterogeneity of transport modes for linear DNA to compensate for the reduced temporal fluctuations. This effect is in fact what is demonstrated by the increase in $\Delta f(t)$ for linear DNA upon crosslinking (Fig 4D, E).

## Discussion

Our collective results indicate that ring DNA employs a wide range of transport modes, each with corresponding distinct conformations, that are not accessible to linear DNA (Fig. 5). These modes are accompanied by chain swelling and suppressed conformational fluctuations that are exacerbated by cytoskeleton crosslinking. As described in the introduction, ring polymers entangled by linear chains have been predicted to assume folded, amoeba-like, and threaded conformations which lead to slower diffusion of rings compared to their linear counterparts. Indeed, ring DNA tracers embedded in solutions of entangled linear DNA have been reported to have diffusion coefficients up to an order of magnitude lower than their linear DNA equivalents (*34, 46, 50*), a phenomenon which has been attributed to threading events. Simulations have also shown that threading leads to anomalous diffusion and swelling of DNA coils (*47–49, 54*) similar to our results. Finally, large fluctuations in the relaxation dynamics of ring DNA embedded in semidilute linear DNA solutions have recently been observed and attributed to threading events (*49*).

Further evidence of threading lies in the biphasic MSDs for rings (Fig. 2A). The lengthscale at which ring DNA MSDs exhibit a shift to more subdiffusive transport (~0.4 μm2) is remarkably close to the squared radius of gyration of rings $R_{G2}$ = 0.42 μm (*46*). Threaded or pinned rings are restricted to move largely perpendicular to the threading filaments and thus can only readily move within a distance ~$R_G$. COM motion of threaded rings over distances larger than $R_G$ can only arise via the slow mechanism of constraint release, which presumably occurs on a timescale comparable to the longest relaxation time of the network (~3 s). As such, this secondary slower phase in COM transport likely arises from the transport mode associated with threaded DNA being partially frozen out at distances >$R_G$. The reduction in ring transport upon crosslinking corroborates this result. Crosslinking of entangled cytoskeleton filaments limits filament diffusion and thus hinders their ability to release their constraints to allow threaded rings to diffuse. As such ring DNA in crosslinked networks can remain threaded for much longer periods of time – even indefinitely – leading to a higher degree of subdiffusion and decreased transport coefficients. This effect is also manifested in the MSD distribution (Fig. S1) that shows a fraction of rings exhibiting nearly zero MSDs. In fact, one may have expected an even larger difference in $K$ and $\alpha$ values upon crosslinking if in entangled networks rings can be released via reptation of filaments, whereas in crosslinked networks they are confined to move <$R_G$ indefinitely. However, because the mesh size of the networks ξ is ~2-3*x* smaller than the DNA coil size (Fig. 1A), it is quite likely that many of the rings are threaded by multiple filaments or become threaded by a new filament before the original threading filament releases its constraint. This phenomenon would prolong the time over which rings remain constrained well beyond the longest relaxation time of the network. This effect also likely contributes to the lack of a long-time plateau in $L_f(t)$ for rings.

The simplified model described above ignores the fact that threaded rings could also move along the backbones of the threading filaments rather than simply perpendicular to them. However, this motion would be confined to the mesh size of the network. If a large fraction of the rings is threaded then we would expect the MSDs to be restricted to $<\xi_2$, particularly for times less than the network relaxation time. This effect is indeed manifested in Fig. 2A in which the MSDs for rings remain $<\xi_2$ over the entire measurement, and one may argue that the curves appear to be asymptoting to $\xi_2$. Conversely, the MSDs for linear DNA surpass $\xi_2$ and display no asymptotic behavior.

The question remains as to why crosslinking facilitates the transport of linear DNA given that it serves to restrict the reptation of the entangling linear cytoskeleton filaments. Further, this increased transport is coupled with larger conformations, at odds with the Newtonian Stokes relationship ($D \sim R_{coil-1}$). We previously showed that linear DNA diffusing in a network of semiflexible actin filaments was more compact and displayed more extreme subdiffusion and lower transport coefficients than when diffusing in a network of more rigid microtubules (19). Further, recent simulations have shown that the slow mobility of large crowders, which results in continuous temporal evolution of the crowding mesh, was required for true anomalous subdiffusion (67). In contrast, rigid constraints (such as crosslinked filaments) resulted in less extreme transient subdiffusion arising from temporary caging of particles in the rigid mesh coupled with hopping to new pockets in the mesh. In a more mobile network, hopping is avoided because the particle motion is coupled to the crowding network motion such that it traverses voids in the mesh by the slow rearrangement of the network. This diffusive mechanism leads to more pronounced subdiffusion and more homogeneous transport than caging and hopping – exactly as we see for linear DNA in entangled versus crosslinked composites. Finally, we previously showed that the more extreme subdiffusion seen in actin networks compared to microtubules was linked with more compact conformations due to more persistent trapping of particles. Likewise, we find that linear DNA assumes more compact conformations in entangled compared to crosslinked networks.

In summary, we have combined single-molecule conformational tracking with ensemble DDM transport analysis to elucidate the transport properties of linear and ring DNA molecules within *in vitro* cytoskeleton networks. We reveal the intriguing role that DNA topology plays in transport and how this role varies with the introduction of crosslinking into cytoskeletal networks (Fig. 5). We find overwhelming evidence of threading of ring DNA by cytoskeleton filaments, resulting in slow anomalous diffusion coupled with a heterogeneous ensemble of transport modes with corresponding swollen conformational states. Further, for ring DNA, crosslinking results in slower and more anomalous diffusion due to cytoskeleton filaments threading the rings and crosslinkers reducing the rate at which rings can be unthreaded via constraint release (68, 69). In contrast, linear DNA displays faster and less subdiffusive transport that is largely homogeneous and coupled to more compact conformations compared to rings, indicating a single diffusive transport mode. Moreover, crosslinking the cytoskeleton network actually leads to faster and less subdiffusive dynamics along with more heterogeneous transport for linear DNA. This unexpected phenomenon likely arises from the increased rigidity of the network, which leads to caging and hopping rather than slow DNA transport coupled to the dynamics of the network.

Our collective results reveal for the first time the critical role that DNA topology plays in cytoskeleton transport and how altering cytoskeleton connectivity can enable a myriad of conformational and transport dynamics of biopolymers across scales. Specifically, we demonstrate the important role that threading could play in intracellular transport of ring DNA. Without crosslinkers, cytoskeleton filaments in cells can reptate to both thread and unthread molecules, resulting in threaded molecules that are largely immobile (i.e. confined by $R_G$ and $\xi$) and unthreaded ones that are more mobile. Even when crosslinkers are present cytoskeleton filaments polymerize and depolymerize in cells, so there are situations where a crosslinked network could form in the presence of ring DNA and thread it, thereby immobilizing it. Beyond the biological implications of our work, our results provide key insights into the poorly understood physics of entangled and crowded ring polymers as well as topological polymer blends important to materials engineering and industrial applications.

## Methods and Materials

**DNA**: Double-stranded 115 kbp DNA is prepared through replication of bacterial artificial chromosomes in *Escherichia coli,* followed by purification and extraction as described previously (*46*). Following purification, supercoiled circular DNA is converted to linear and ring (relaxed circular) topologies through treatment with MluI and topoisomerase-I (New England Biolabs), respectively (*57*). In all experiments DNA is fluorescent-labeled with YOYO-1 (Thermofisher) at a basepair:dye ratio of 4:1 (*58*).

**Cytoskeleton proteins:** Composite networks of either entangled or crosslinked actin and microtubules are prepared using previously described protocols (*68*). Briefly, a 1:1 molar ratio of porcine brain tubulin dimers and rabbit skeletal actin monomers (Cytoskeleton) are resuspended to a final protein concentration of 5.8 µM in an aqueous buffer consisting of 100 mM PIPES, 2 mM $MgCl_2$, 2 mM EGTA, 1 mM ATP, 1 mM GTP, and 5 µM Taxol (*15*). Final solutions are pipetted into capillary tubing, sealed with epoxy, and incubated for 30 minutes at $37_o$C to polymerize proteins and form composite networks. For crosslinked composites, we pre-assemble biotin-NeutrAvidin crosslinker complexes as described previously (*68*) and add them to the protein solutions at a crosslinker:protein molar ratio of $R_{CP} = 0.02$ before incubation (*68*). Both networks, fully characterized in Refs. 15 and 68, consist of randomly oriented filaments with minimal bundling and no phase separation between proteins (*15, 68*). The composite mesh size is $\xi \approx 0.81$ µm (*15*). In the crosslinked composite, if we assume that all crosslinkers are incorporated in the network, the length between crosslinkers along an actin filament would be $l_{c,a} = \frac{1}{2}l_{mon} \times R_{C\text{-}1} = 135$ nm where $l_{mon} = 2.7$ nm is the length that each actin monomer adds to an actin filament. Similarly, the length between crosslinkers along a microtubule would be $l_{c,m} = \frac{1}{2}(l_{ring}/13) \times R_{C\text{-}P\text{-}1} = 30$ nm where every 13 tubulin dimers adds $l_{ring} = 7.8$ nm in length to the microtubule. These length scales are smaller than $\xi$, which would suggest bundling of filaments that would in turn increase the mesh size. However, we find no evidence of these effects in composites. We can therefore assume that there is a fraction of crosslinkers that remain free in solution and not incorporated into the network which would increase $l_{c,a}$ and $l_{c,m}$ and thus limit bundling. While we cannot determine this fraction, given that $l_c$ is substantially smaller than $\xi$, we assume that every filament entanglement is permanently linked in the crosslinked composite.

**Sample Preparation:** For all experiments, YOYO-labeled linear or ring DNA is added to the protein solution prior to loading into capillary tubing at concentrations of 0.25 µg ml-1 or 26 µg ml-1 for single-molecule or DDM measurements, respectively. Glucose (0.9 mg ml-1), glucose oxidase (0.86 mg ml-1), and catalase (0.14 mg ml-1) are also added to inhibit photobleaching; 0.05% Tween is added to prevent surface interactions.

**Imaging and Analysis:** DNA molecules within composites are imaged using a home-built lightsheet microscope with a 10x 0.25 NA excitation objective, a 20x 0.5 NA imaging objective, and an Andor Zyla 4.2 CMOS camera.

*Single-molecule conformational tracking:* For each sample, 45 videos displaying ~10 DNA molecules per frame are recorded at 10 fps for 500 frames. All data presented are for an ensemble of ~1000 molecules from 2 different samples, each tracked for a minimum of 2.5 s. Custom-written software (Python) is used to track the center-of-mass (COM) positions $(x, y)$ as well as the lengths of the major axis ($R_{max}$) and minor axis ($R_{min}$) of each molecule in each frame. From COM positions, we compute the mean-squared displacement $MSD = \frac{1}{2}(<(\Delta x)_2> + <(\Delta y)_2>)$ and corresponding transport coefficients and scaling exponents via $MSD = Kt\alpha$ (Figs. 1b, 2, S1). For normal Brownian motion, $\alpha = 1$ and $K$ is equivalent to twice the diffusion coefficient $D$. From the major and minor axis length measurements, we calculate an effective coil size $R_{coil} = [\frac{1}{2}(R_{max2}+R_{min2})]^{\frac{1}{2}}$ (Figs. 1B, 3) (*19*). Finally, we characterize the time-dependent conformational fluctuations of single molecules by calculating the fractional fluctuation length $L_f(t) = <|R_{max}(t) - R_{max}(0)|>/<R_{max}>$ for all lag times $t$. $L_f(t)$ quantifies the timescale and fractional lengthscale over which single molecules fluctuate between different conformational states. These analysis methods, depicted in Figure 1, have been described and validated previously (*19–21, 35*).

*DDM*: For each sample, 8 videos with a 256 x 1280 px (49.6 x 248.3 μm₂) field of view are recorded at 18 fps for 5000 frames at different regions within the sample. For DDM analysis (described in Ref. 62), videos are then split into 256 x 256 px (49.6 x 49.6 μm₂) ROIs (*19, 62–64*). Each ROI is analyzed individually and averaged together post-analysis (Fig. 1C). A 2D Fourier transform is taken of the difference between images separated by time lags of 0.05 to 166.55 s (Fig. 1C, 2). Due to the dynamics being isotropic, the 2D Fourier transform is radially averaged for all lag times $t$, resulting in the DDM matrix $D(q,t)$ where $q$ is the magnitude of the wave vector. The DDM matrix can be fit to $D(q,t) = A(q)[1-f(q,t)]+B(q)$ where $f(q,t)$ is the intermediate scattering function. We use a stretched exponential for the ISF (Fig. S2). While such fits for each $q$ do not follow the data over all $t$ for ring DNA, they do allow us to extract the parameters $A(q)$ and $B(q)$. With the measured $D(q,t)$ and extracted $A(q)$ and $B(q)$, we plot the ISF, $f(q,t)$, for a particular wave vector in order to compare the rate at which the ISF decays (Fig. 1C, 3).

### Acknowledgements

We thank S. Ricketts for her work optimizing and characterizing actin-microtubule polymerization and crosslinking protocols. We thank Gina Aguirre for her help in analyzing data. **Funding** This work was supported by NIH-NIGMS Award # R15GM123420 to R.M.R.A. and R.M., AFOSR Award # FA9550-17-1-0249 to R.M.R.A., ACS PRF Award # 57326-UNI10 to R.M., and NSF Award CBET-1603925 to R.M.R.A. **Author contributions:** R.M.R.A and R.M. conceived the project, guided experiments, analyzed and interpreted data, and wrote the manuscript. D.W. and K.R. performed experiments, analyzed data, and wrote the manuscript. J.G. analyzed data. **Competing interests:** the authors have no competing interests to declare. **Data and materials availability**: All data needed to evaluate the conclusions of the paper are present in the paper and/or the Supplementary Materials. Additional data related to this paper may be requested from the authors.

**Figures:**

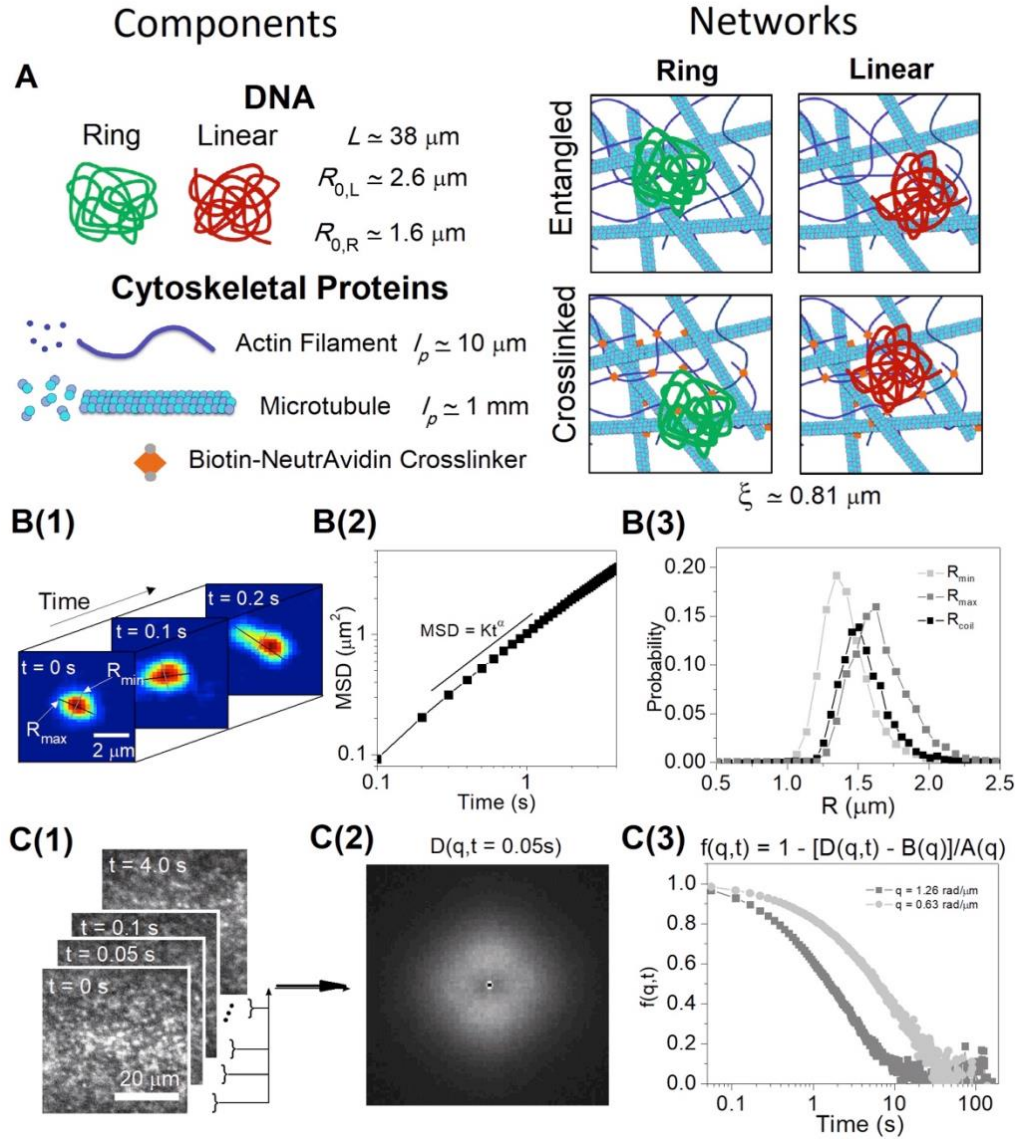

**Figure 1. Experimental approach to elucidate the effect of DNA topology on the transport and conformational properties of DNA diffusing through model cytoskeleton composites.** (A) Cartoon of fluorescent-labeled 115 kbp ring and linear DNA molecules embedded in composite networks of actin and microtubules that are either entangled or crosslinked by biotin-NeutrAvidin crosslinkers. $L$ is the DNA contour length, $R_0$ is the topology-dependent mean end-to-end length of the DNA coils, $l_p$ is persistence length, and $\xi$ is the composite mesh size. Not drawn to scale. (B) Single-molecule analysis (1) tracks the center of mass (COM) position and the lengths of the major and minor axes ($R_{max}$ and $R_{min}$) of each DNA molecule for every frame of the time-series to quantify the transport and conformational dynamics of individual DNA molecules. From (1), the COM mean squared displacement (MSD) (2) and probability distributions of $R_{max}$, $R_{min}$, and $R_{coil} = [\frac{1}{2}(R_{max2}+R_{min2})]_{1/2}$ (3) are computed. (C) From the differences in images separated by a given lag time (1), DDM analysis computes the matrix $D(q,t)$ where $q$ is the magnitude of the wave vector (2). The intermediate scattering functions $f(q,t)$ versus lag time for each spatial frequency $q$ describes the ensemble dynamics (3).

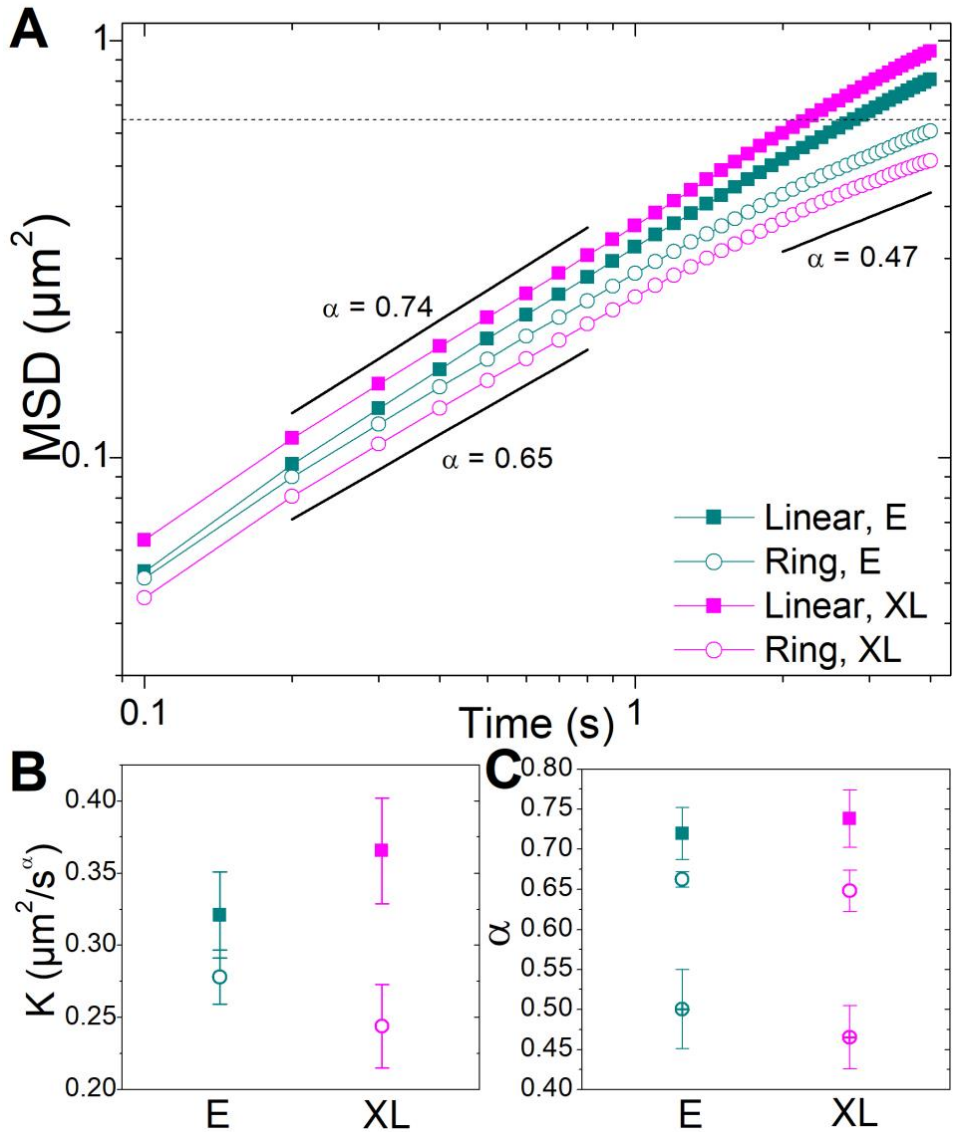

**Figure 2. Ring DNA in cytoskeleton composites exhibits unique two-phase subdiffusion distinct from linear DNA and amplified by cytoskeleton crosslinking.** (A) Mean-squared displacements (MSD) versus time for ring (open circles) and linear (closed squares) DNA in entangled (cyan, E) and crosslinked (magenta, XL) actin-microtubule networks. Horizontal dashed line denotes where $MSD = \xi_2$. Black lines represent power-law scaling with exponents listed. Fits of the MSDs to the power-law relation $MSD = Kt_\alpha$ yield transport coefficients $K$ (B) and scaling exponents $\alpha$ (C) for linear (closed squares) and ring (open and crossed circles). MSDs for linear DNA obey a single power-law over the entire measurement time (squares in (C)), while rings exhibit a second slower phase with lower $\alpha$ values (crossed circles in (C)) starting at ~0.4 µm2. (C) Linear DNA exponents are determined from fits over $t = 0.1 - 4$ s (closed squares), while ring DNA exhibits two different exponents with values determined from fits over $t = 0.1 - 2$ s (open circles) and $t = 2 - 4$ s (crossed circles). As shown, linear DNA exhibits faster transport and less subdiffusion in crosslinked compared to entangled networks, while crosslinking has the opposite effect on ring DNA.

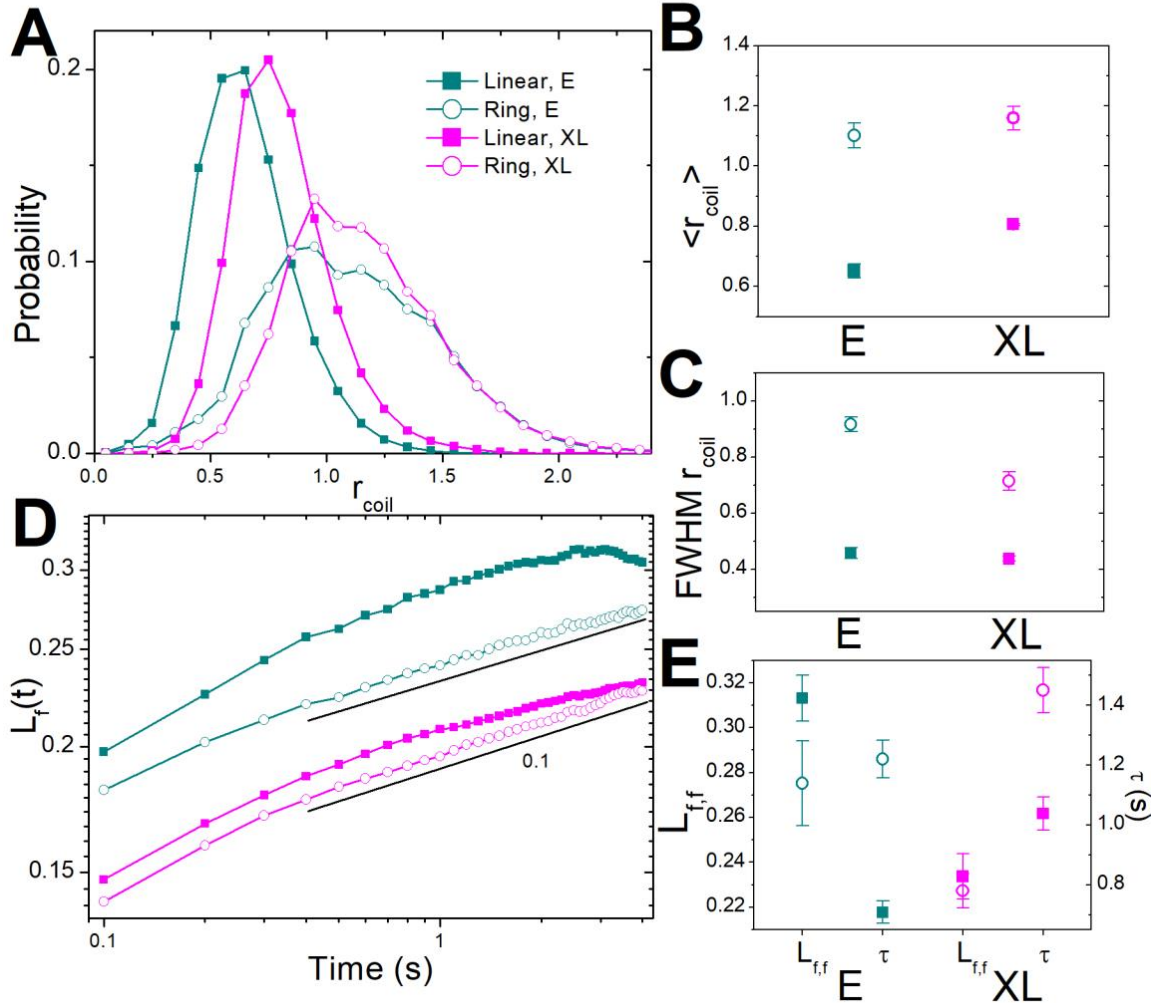

**Figure 3. Both DNA topology and cytoskeleton crosslinking impact the conformational dynamics of ring and linear DNA in cytoskeleton composites.** (A) Probability distributions of the coil sizes $R_{coil}$ for every frame of every molecule. $R_{coil}$ is rescaled by the expected dilute-limit end-to-end distance $R_0$, which we denote as $r_{coil}$. Distributions show compaction of linear DNA (squares, distribution centered at $< 1$) from normal $R_0$ values, while ring DNA (open circles) swells and accesses a broader range of coil sizes (distribution centered at >1, broader than linear DNA distributions). (B) Mean rescaled coil sizes $< r_{coil}>$ quantify the swelling of rings (circles) and compaction of linear DNA (squares) in entangled (E, cyan) and crosslinked (XL, magenta) networks. (C) Full-width-half-maximum (FWHM) of $r_{coil}$ distributions shown in A, displaying the topology-dependent range of conformational states accessed by DNA. (D) Fractional fluctuation length $L_f(t) = <|R_{max}(0) - R_{max}(t)|>/<R_{max}>$ for linear and ring DNA with black lines denoting power-law scaling with exponent listed. Linear DNA fluctuates more quickly and over a larger range than ring DNA, approaching steady-state values compared to the slow power-law rise of ring DNA. (E) The final fractional fluctuation length, termed $L_{f,f}$ plotted alongside the time $\tau$ at which molecules reach 90% of $L_{f,f}$. As shown, ring DNA fluctuates more slowly *and* over a smaller range than linear DNA in both entangled and crosslinked networks.

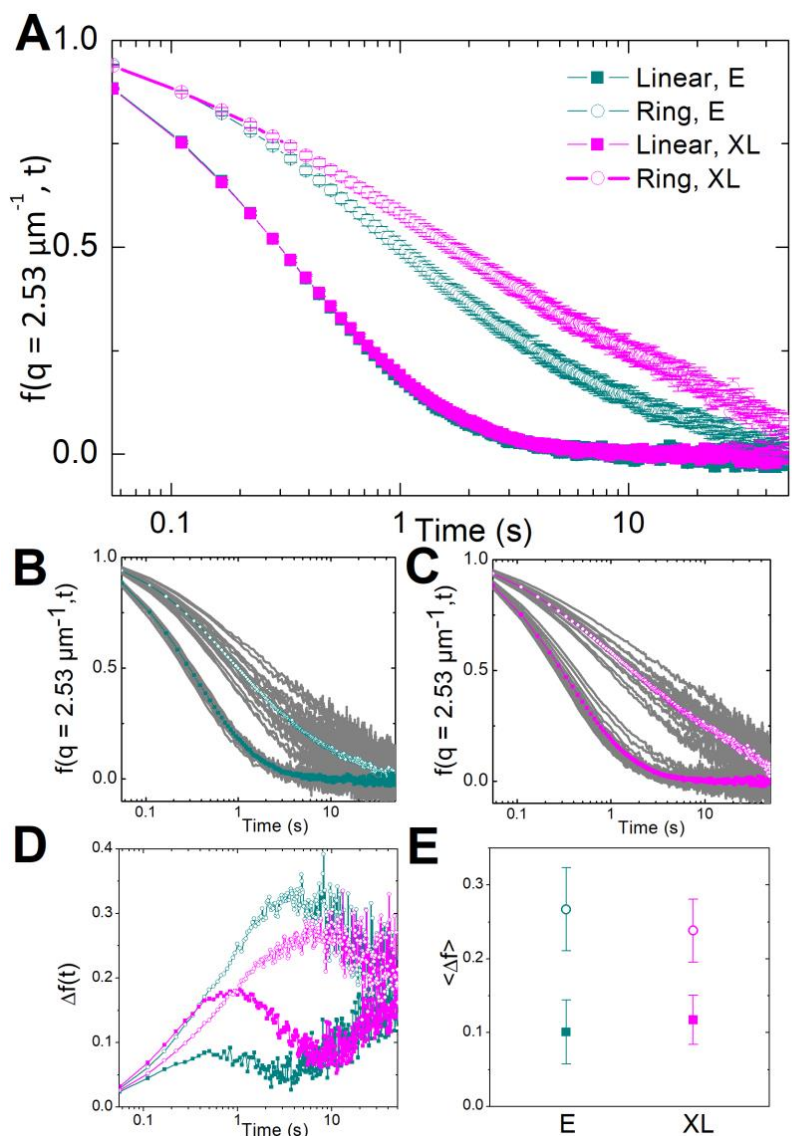

**Figure 4. DDM reveals heterogeneous slow transport of ring DNA with unique sensitivity to crosslinking.** (A) Average intermediate scattering function (ISF) $f(q,t)$ with $q = 2.53$ rad/μm for ring (circles) and linear (squares) DNA in entangled (cyan) and crosslinked (magenta) cytoskeleton composites. Displayed curves are averages over 20 regions of interest. ISFs for linear DNA decay much faster than for ring DNA and exhibit expected exponential decay not seen for rings. While crosslinking slows the decay for rings it has a negligible effect on the linear DNA ISF. (B, C) All individual ISFs (grey) comprising the average ISF (color coded) for linear and ring DNA in entangled (B) and crosslinked (C) composites. The substantial spread in ISF curves for ring DNA and the slow decay to zero – both features absent for linear DNA – indicate heterogeneous or multi-mode transport and anomalous slow diffusion, respectively. ISFs for crosslinked networks show a slightly smaller spread for ring topologies while the spread for linear topologies is slightly larger. (D) That spread in ISFs is quantified by taking the difference between the maximum and minimum values of $f(q,t)$, $\Delta f(q,t)$, among the multiple regions of interest at $q = 2.53$ rad/μm. (E) The average $\Delta f(q,t)$ over the range of time lags is greater with ring than with linear DNA for both networks. For rings, moving from an entangled to a crosslinked network decreases the spread. Conversely, for linear DNA, the spread increases slightly upon crosslinking.

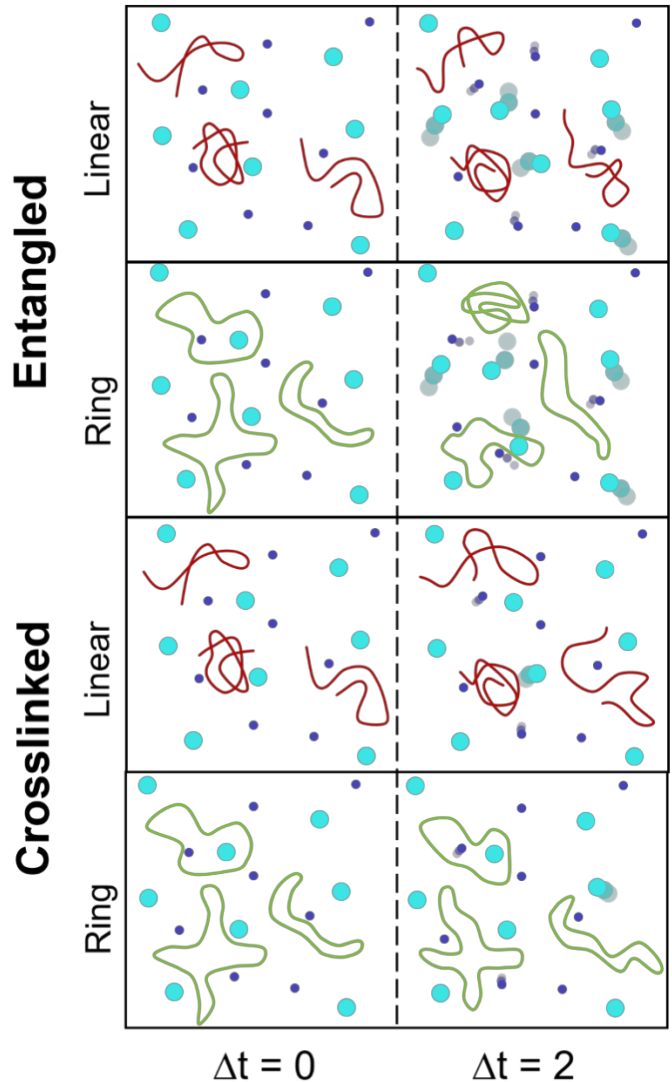

**Figure 5. Ring DNA molecules in cytoskeleton networks adopt multiple modes of transport that are not accessible to linear DNA and are impacted by cytoskeleton crosslinking.** Cartoon of ring (green) and linear (red) DNA diffusing through entangled and crosslinked networks of actin (purple) and microtubules (blue). Timescale is arbitrary and cartoons not drawn to scale. Each panel is a depiction of a slice in the *xy*-plane with DNA aligned in the plane and actin and microtubule constraints oriented along *z*. In right-hand panels (*Δt*=2), lighter shaded circles denote the two previous positions (*Δt* = 0, 1) of the corresponding constraints. Within entangled networks, linear DNA can reptate through the network, while ring DNA adopts branched, folded, or threaded conformations. Reptation of the entangled cytoskeleton filaments allows threaded rings to become unthreaded via constraint release of the threading filaments. Crosslinking suppresses the mobility of cytoskeleton filaments which can cause rings to become permanently threaded, slowing their transport, while at the same time increasing the mobility of linear DNA as described in text (Fig. 2). Cartoons are not drawn to scale.

**Topology-dependent anomalous dynamics of ring and linear DNA are sensitive to cytoskeleton crosslinking**

Devynn M. Wulstein[+], Kathryn E. Regan[+], Jonathan Garamella, Ryan J. McGorty[#], Rae M. Robertson-Anderson[#*]

*Department of Physics & Biophysics, University of San Diego, San Diego CA 92110*

[+]equal contributions
[#]equal contributions
*corresponding author, randerson@sandiego.edu

**Supplementary Materials**

Table S1: Quantities derived from single-molecule conformational dynamics analysis

Figure S1: Distribution of individual MSDs from SMCT analysis

Figure S2: DDM Fits

| | **Linear, E** | **Linear, XL** | **Ring, E** | **Ring, XL** |
|---|---|---|---|---|
| **$<r_{coil}>$** | $0.65 \pm 0.02$ | $0.806 \pm 0.003$ | $1.10 \pm 0.04$ | $1.16 \pm 0.04$ |
| **FWHM $r_{coil}$** | $0.46 \pm 0.02$ | $0.438 \pm 0.009$ | $0.92 \pm 0.03$ | $0.71 \pm 0.03$ |
| **$L_{ff}$** | $0.31 \pm 0.01$ | $0.23 \pm 0.01$ | $0.28 \pm 0.02$ | $0.227 \pm 0.008$ |
| **$\tau$ (s)** | $0.71 \pm 0.04$ | $1.04 \pm 0.05$ | $1.22 \pm 0.06$ | $1.45 \pm 0.08$ |

**Table S1: Quantities derived from single-molecule conformational dynamics analysis.** Averages and standard errors for quantities plotted in Fig 3B,C,E: $<r_{coil}>$, FWHM $r_{coil}$, $L_{ff}$ and $\tau$.

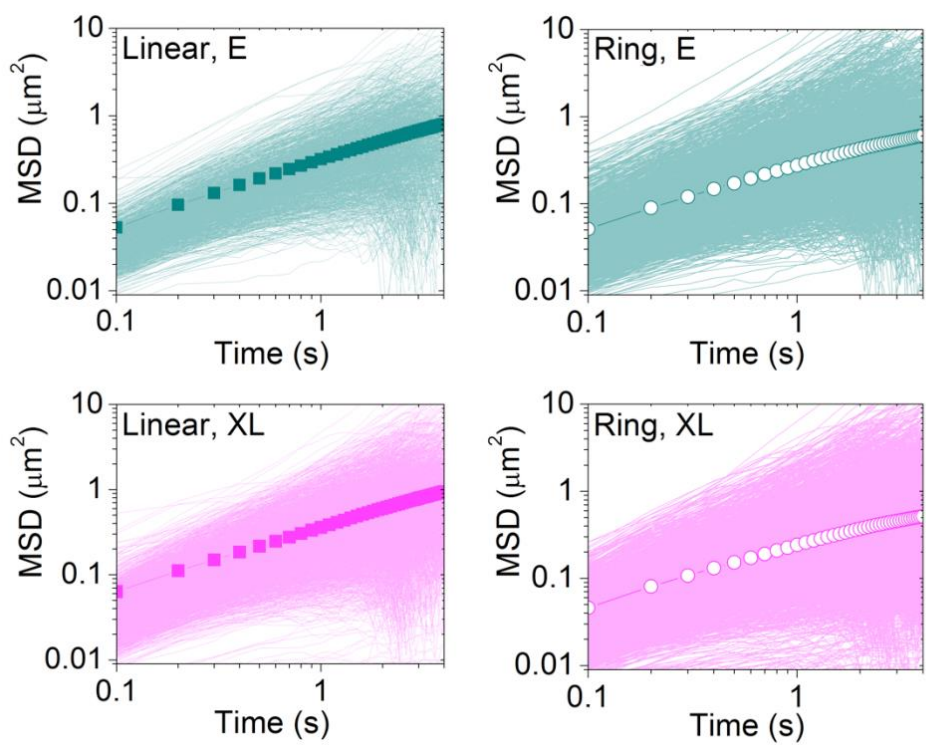

**Figure S1: Distribution of individual MSDs from SMCT analysis.** Mean squared displacements for all ~1000 molecules in each ensemble, shown as light cyan and light magenta lines, for ring and linear DNA in entangled (E, cyan) and crosslinked (XL, magenta) cytoskeleton networks. In each plot the ensemble-averaged MSD determined from the distribution shown is overlaid using symbols as in Figure 2. The DNA topology and network type are listed at the top of each plot.

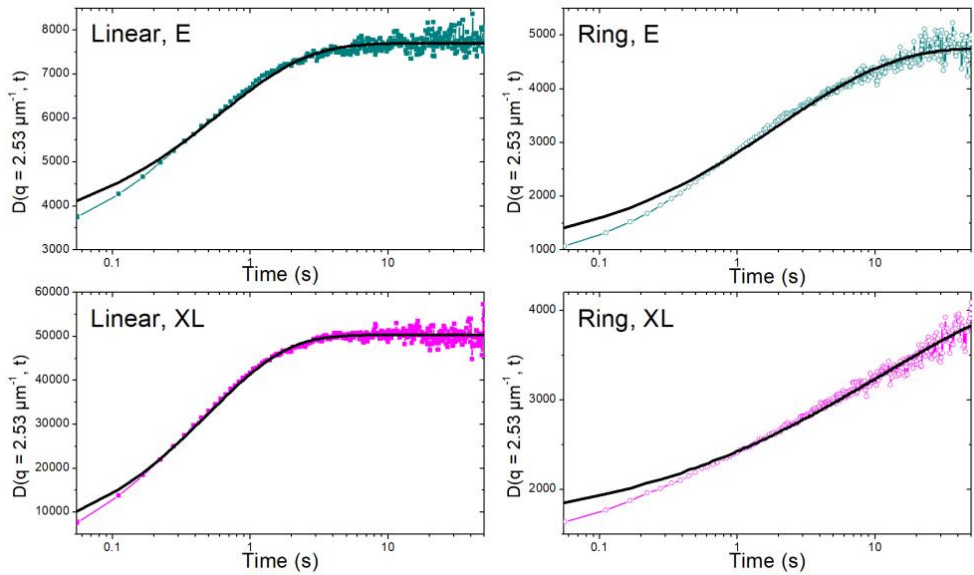

**Figure S2: Fits to *D(q,t)* from DDM analysis.** We plot the DDM matrix *D(q,t)* for *q* = 2.53 μm⁻¹ as a function of lag time for ring and linear DNA in entangled (E, cyan) and crosslinked (XL, magenta) cytoskeleton networks. For each combination, we plot the fit to $D(q,t) = A(q)[1-f(q,t)]+B(q)$ as a solid black line. For the intermediate scattering function, *f(q,t)*, we use a stretched exponential: $f(q,t) = \exp[-t/\tau]^\beta$. From these fits we extract *f(q,t)* curves which are plotted in Figure 4.